\documentclass[aps,prb,twocolumn,groupedaddress,showpacs]{revtex4}
\usepackage[dvips]{graphicx}

\begin{document}

%\preprint{}

\title{Non-equilibrium breakdown of quantum Hall state in graphene}


\author{Vibhor~Singh}
\author{Mandar~M.~Deshmukh}
\email[]{deshmukh@tifr.res.in}
%\thanks{}

\affiliation{Department of Condensed Matter Physics and Materials
Science, Tata Institute of Fundamental Research, Homi Bhabha Road,
Mumbai, 400005 India }


\date{\today}

\begin{abstract}
In this report we experimentally probe the non-equilibrium breakdown
of the quantum Hall state in monolayer graphene by injecting a high
current density ($\sim$1A/m). The measured critical currents for
dissipationless transport in the vicinity of integer filling factors
show a dependence on filling factor. The breakdown can be understood
in terms of inter Landau level (LL) scattering resulting from mixing
of wavefunctions of different LLs. To further study the effect of
transverse electric field, we measured the transverse resistance
between the $\nu=2$ to $\nu=6$ plateau transition for different bias
currents and observed an invariant point.

\end{abstract}

%\keywords{}

\maketitle

%\section{}
%\subsection{}
%\subsubsection{}

%************************************************
The quantum Hall effect (QHE) \cite{qhbook1} has been studied
extensively in 2D systems and its equilibrium electron-transport
properties are understood to a large extent. The breakdown of the
QHE under non-equilibrium conditions due to a high current density
has been studied to understand its microscopic origin\cite{nachtwei,
Ebert}. There has been a considerable debate in the literature
regarding the details of the mechanism of QHE breakdown. The
proposed mechanisms include electron heating \cite{electronheating},
electron-phonon scattering \cite{Streda, electron-phononscattering},
inter and intra Landau level (LL) scattering\cite{EavesAndSheard,
inter-intra-landauleveltunneling1}, percolation of incompressible
regions \cite{luryi} and the existence of compressible regions in
the bulk \cite{thouless}.
\begin{figure}
\includegraphics[width=65mm]{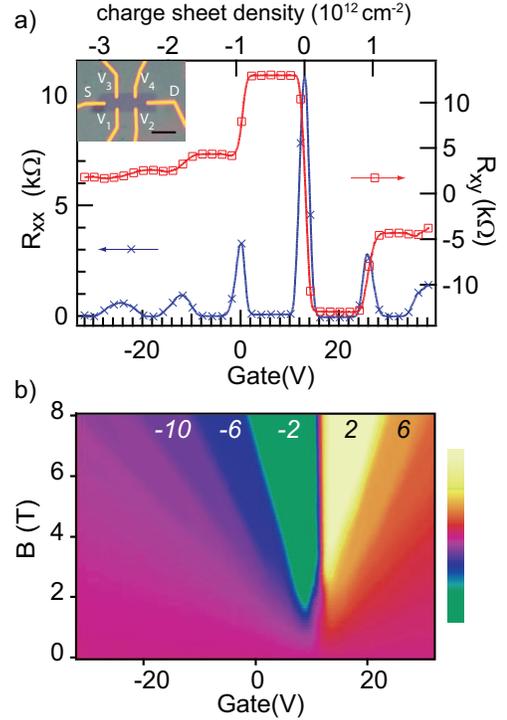}
\caption{\label{fig:figure1} (color online) a) Line plot of the
longitudinal resistance ($R_{xx}$) and transverse resistance
($R_{xy}$) for a monolayer graphene device at 300~mK and 9~T. The
inset shows an optical microscope image. The scale bar corresponds
to 6 $\mu m$. Probes $S$ and $D$ were used to current bias the
device. By using a lock-in technique, probe pairs $V_1-V_2$ and
$V_1-V_3$ were used to measure $R_{xx}$ and $R_{xy}$ respectively.
50 nA of AC current at 181 Hz was used for these measurements. b)
Colorscale plot of $R_{xy}$, to show the plateaus of varying
magnitude clearly, as a function of magnetic field at 300~mK.}
\end{figure}

Recently QHE has also been observed in graphene\cite{firstgraphene}
and studied extensively\cite{grapheneRMP}. In this paper we probe
the breakdown of the QHE in graphene by injecting a high current
density ($\sim$~1A/m); this results in a large local electric field
in the system. The unique band structure of graphene near the Fermi
energy ($E=\pm c\hbar |k|$, where $c\approx10^6$ ms$^{-1}$ is the
Fermi velocity) gives rise to a `relativistic' QHE. In a magnetic
field perpendicular to its plane, the energy spectrum of graphene
splits into unequally spaced LLs and is given by
$E_n=sgn(n)\sqrt{(2\hbar c^2 e B |n|)}$, where $n$ is the LL index.
When the Fermi level lies between two LLs, the longitudinal
resistance ($R_{xx}$) vanishes and the transverse resistance
($R_{xy}$) gets quantized to $\frac{h}{(4|n|+2)e^2}$. Further, the
presence of transverse electric field mixes the electron and hole
wavefunctions and modifies the energy
spectrum\cite{Baskaran,nietoQHE,KramerGraphene}, which is given by
$E_n=sgn(n)\sqrt{(2\hbar c^2 e B |n|)} (1-\beta^2)^{3/4} -\hbar c
\beta k_{\bot}$, where $\beta=E/(cB)$, $E$ is the electric field
orthogonal to $B$, and $k_{\bot}$ is the wave vector in the
direction perpendicular to $E$ and $B$.

\begin{figure}
\includegraphics[width=70mm]{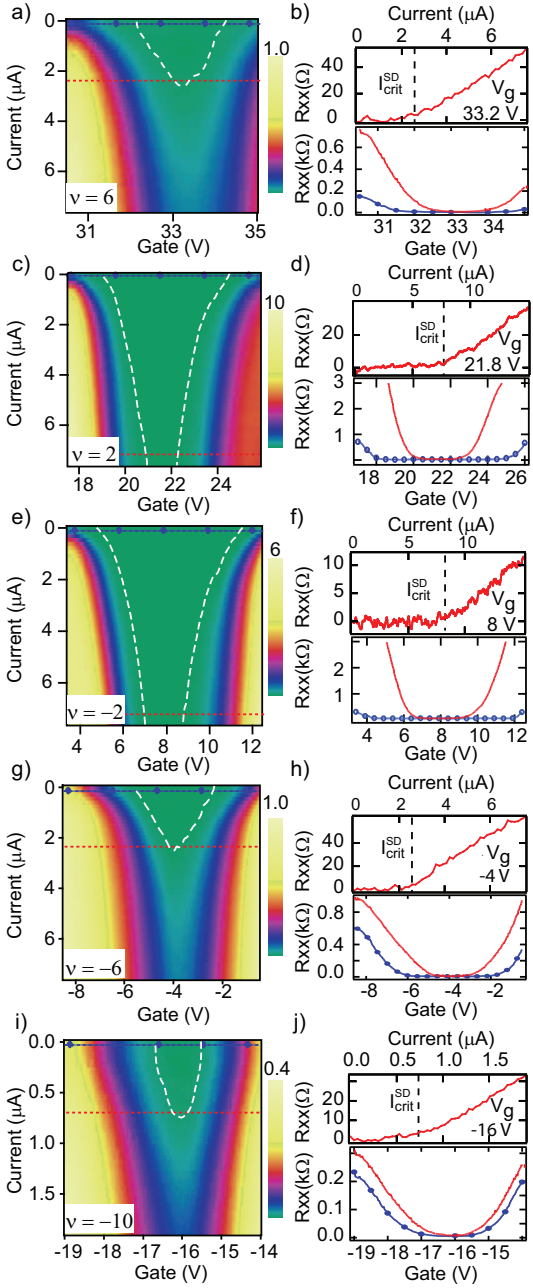}
\caption{\label{fig:figure2} (color online) Critical current
measurements in the vicinity of integer filling factors at 9~T and
300~mK. a, c, e, g and i show the colorscale plot of $R_{xx}$ as a
function of $I_{DC}^{SD}$ and $V_g$ near filling factors 6, 2, -2,
-6, and -10 respectively. Color bars indicate the resistance in
units of k$\Omega$. The white dotted lines on the colorscale plot
mark the dissipationless region. The line plots in b, d, f, h and j
show slices along the current axis at the gate voltages shown in the
figure (top) and two slices (bottom) each for the equilibrium
(labeled with solid circles) and non-equilibrium (solid line)
biasing conditions. The position of these slices is marked in
adjoining colorscale plots (marked a, c, e, g and i ).}
\end{figure}

 The motivation for exploring the breakdown of QHE in graphene
is twofold -- first, the QHE in graphene is very different from the
QHE in a 2DEG system. The LL energy spectrum of 2DEG is equispaced
unlike that in graphene. The energy scale set by the cyclotron gap
$(\Delta E_{\nu})$ in graphene at $B = $10~T, is much higher
($\sim$1300~K) than its value for a 2DEG ($\sim$20~K) at the same
magnetic field\cite{grapheneRMP}. The mechanism of breakdown in
graphene could be the inter-LL scattering due to wavefunction-mixing
or possibly entirely different if the lengthscale for variation of
the local electric field due to defects is comparable to the
magnetic length. In such a situation, ($\beta\ge 1)$, a ``collapse"
of the LL is possible before a longer lengthscale breakdown of QHE
\cite{Baskaran}. Second, graphene shows room temperature QHE
\cite{RoomQHE} at high magnetic field, therefore understanding the
breakdown mechanism can also be useful for metrological resistance
standards \cite{GeimRecentAPL}. In addition, the presence of back
gate in our devices allows us to change the Fermi level. We can
hence probe the QH breakdown away from the integer filling factors
without changing the energy spectrum.  With these motivations in
mind we have probed the breakdown of QHE near the filling factors
$(sgn(n)(4|n|+2))$ \cite{noteaboutnu,firstgraphene}, $\nu= -10,-6,-2,2,6$
\cite{absenceofnu10}. To better understand the role of high current
densities we have modeled the effect of high current using a
current-injection model\cite{KramerGraphene}; this allows us to
explain the experimentally observed transition from neighboring
filling factors in terms of an invariant point. We also provide
evidence to show that these experimental observations cannot be
explained purely on the basis of local electronic heating.

To fabricate monolayer graphene devices in a Hall bar geometry, we
have followed the mechanical exfoliation
technique\cite{firstgraphene,grapheneRMP} on degenerately-doped
silicon substrates coated with 300 nm thick SiO$_2$. We optically
locate the flakes of graphene and pattern electrodes onto them using
electron beam lithography. The electrodes are fabricated by
depositing 10~nm Cr and 50~nm of Au by thermal evaporation. The
degenerately-doped silicon substrate serves as a back-gate to tune
the density of carriers by applying a voltage $V_g$. The inset of
Fig.\ref{fig:figure1}a shows an optical image of a Hall bar device.
In Fig.\ref{fig:figure1}a we plot the longitudinal resistance
($R_{xx}$) and transverse resistance ($R_{xy}$) at $T$ = 300~ mK and
$B =$ 9 T. Filling factors, unique to monolayer graphene ($\nu=\pm2,
\nu=\pm6$) are clearly seen for both types of carriers in
Fig.\ref{fig:figure1}a. The mobility of the device shown in the
inset is measured to be $\sim$11000 cm$^2$(Vs)$^{-1}$ for both types
of carriers at 300 mK; by using the semi-classical relation for mean
free path $(l)$,\cite{transportSDS} the measured mobility gives
$l=70$~nm for carrier density $3\times10^{11}$ cm$^{-2}$. In
Fig.\ref{fig:figure1}b we plot the evolution of $R_{xy}$ as a
function of $V_g$ and magnetic field $B$. The plateaus in $R_{xy}$
corresponding to $\nu=$ $\pm$ 2, $\pm$ 6 and $-10$ are clearly seen.
The Dirac peak for the device is shifted to about 13~V due to
unintentional doping, which corresponds to a charge inhomogeneity of
$6\times10^{11}$ cm$^{-2}$ \cite{transportSDS}. From the
two probe resistance measurements at $B =$ 8 T, we find that contact
resistance is smaller than 700 $\Omega$ in the QH regime.

To probe the breakdown of QHE, we biased the source-drain probes of
our device with DC current ($I_{DC}^{SD}$) along with a small AC
current (50 nA) in the minima of $R_{xx}$ corresponding to filling
factors $\nu= \pm 2, \pm 6$ and $ -10$ \cite{absenceofnu10} at fixed
magnetic field. The AC current remains fixed and $I_{DC}^{SD}$ is
then varied as a function of $V_g$ in the vicinity of integer $\nu$.
The AC signals between the voltage probes $V_1$ and $V_2$ and the
voltage probes $V_1$ and $V_3$ were monitored with a lock-in
amplifier to record the values of $R_{xx}$ and $R_{xy}$
respectively. Fig.\ref{fig:figure2} shows the evolution of the
$R_{xx}$ minima as function of $I_{DC}^{SD}$ for different filling
factors. The line plots show slices of the data in equilibrium and
non-equilibrium biasing conditions. In order to interpret the
breakdown from the measured experimental data we define a critical
current ($I_{crit}^{SD}$) as the linearly extrapolated value of
$I_{DC}^{SD}$ at zero dissipation\cite{kirtley1}.

We point out the qualitative features of our data -- first, with the
increase of $I_{DC}^{SD}$, the width of the dissipationless region
reduces, eventually leading to the breakdown. Second, the critical
current is $\nu$ dependent. Third, on either side of the integer
filling factor the boundary of dissipation evolves asymmetrically.
In Fig.\ref{fig:figure2}a, c, e, g and i for $\nu=6,2,-2,-6$ and
$-10$ respectively we see a non-linear evolution of the boundary of
dissipation as a function of $I_{DC}^{SD}$ and $V_g$.

Before discussing details of the data, we address the possible
concerns about local heating of the sample that could occur in these
studies. We have done control experiment to confirm that heating is
not responsible for the key experimental observations; by comparing
change of resistance with temperature and due to current. We
injected a large DC current through the voltage probes $V_1$ and
$V_2$ of the Hall bar geometry device to intentionally heat it
locally while we simultaneously measured two probe resistance-gate
voltage characteristic of the opposite two probes ($V_3$ and $V_4$)
using a lock-in technique. Then, we observed the evolution of two
probe resistance-gate voltage characteristic of the same pair ($V_3$
and $V_4$) with temperature while the injected DC current was set to
zero. By comparing these two data sets (change of resistance with
temperature and due to current), we could estimate that the
temperature of the device does not increase beyond ~3 K while
injecting currents as high as 10 $\mu$A. We have also seen an
overall asymmetry in the evolution of $R_{xx}$ with the sign of
injected current close to integer filling factor. Pure thermal
effects cannot explain this asymmetry. Additionally, cyclotron gaps
in graphene are large and thermal effects cannot completely suppress
the electric field induced effect. Also, superior thermal
conductivity of graphene\cite{grapheneheatconduction} is likely to
suppress any local thermal hot-spot formation.

\begin{figure}
\includegraphics[width=65mm]{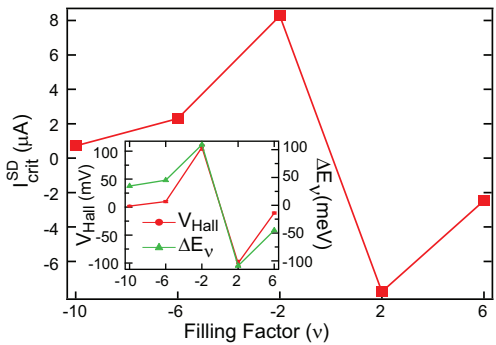}
\caption{\label{fig:figure2b} (color online) Plot of critical
current($I_{crit}^{SD}$) for different filling factors at $T$ = 300
mK and $B$ = 9 T. The inset shows the plot of Hall voltage developed
at breakdown ($V_{Hall}$, labeled with circles) and the cyclotron
gaps ($\Delta E_{\nu}$) (solid triangles) plotted on right axis as a
function of $\nu$.}
\end{figure}

In order to understand the mechanism of the breakdown we examine the
dependance of $I_{crit}^{SD}$ on $\nu$. Fig.\ref{fig:figure2b} shows
the plot of $I_{crit}^{SD}$ for various filling factors $\nu$
indicating that $I_{crit}^{SD}$ decreases with $|\nu|$. The inset
shows the plot of two relevant quantities, Hall voltage,
$V_{Hall}=I_{crit}^{SD}\times \frac{h}{\nu e^2}$, and $\Delta
E_{\nu}$ as a function of $\nu$. There is a correlation between
$V_{Hall}$ and $\Delta E_{\nu}$, which can be explained by
considering inter LL scattering. % here the language was changed to make the connection explicit
The origin of the inter-LL scattering is likely to be the strong
local electric field that mixes the electron and hole wavefunctions
\cite{Baskaran,nietoQHE,KramerGraphene} providing a finite rate for
inelastic transitions. The energy for the transitions is provided by
the transverse electric field parallel to the electron trajectory.
This results in inelastic scattering between LLs leading to a
breakdown of the dissipationless QH state \cite{EavesAndSheard}.
Similar inter-LL scattering mechanisms have been used to explain the
breakdown of the QHE in a 2DEG, in samples of width less than 10
$\mu$m and moderate mobility\cite{nachtwei,EavesAndSheard,
kirtley1}. The electric field for inter-LL ($E_{LL}$) scattering can
be estimated to be that field where the quasiparticle can pick up an
energy corresponding to the LL separation within few cyclotron radii
($r_c$) \emph{i.e.} $e E_{LL}\sim \Delta E_{\nu}/r_c \approx 10^6$
V/m. This is much higher than the experimentally observed electric
fields. However, Martin \emph{et al.} \cite{grapheneSET} found a
much shorter lengthscale associated with the charge inhomogeneity
($\sim$150~nm). The presence of a charge inhomogeneity
\cite{grapheneSET} leads to a strong local electric field and thus
can reduce the threshold for the breakdown due to inter-LL
scattering.

For $\nu=\pm 2$, $V_{Hall}$ matches quite well with $\Delta
E_{\nu=\pm 2}$, which indicates that the $n=0$ LL width is small.
However for $\nu=\pm 6,-10$, $V_{Hall}$ is smaller than the
corresponding cyclotron gap. This deviation for $\nu=\pm 6,-10$ can
be explained by considering disorder-induced broadening of
$n=\pm1,-2$ LLs. The difference between $\Delta E_{\nu}$ and
$eV_{Hall}$ is approximately $\sim$35meV ($\sim$415K). These
observations are consistent with the experiments measuring quantum
Hall activation gap \cite{activationGap}, which have also revealed
similar width of these LLs in samples of similar mobilities.
Additionally, the difference between $V_{Hall}$ and $\Delta E_{\nu}$
can also be attributed to inhomogeneous charge distribution.
Considering inhomogeneous charge distribution, the critical current
is predicted to be filling factor and length scale
dependent\cite{Baskaran}. However, the $n=0$ level remains protected
from the local electric field
fluctuations\cite{Baskaran,AtiyahSingerIndex}. The correlation
between $V_{Hall}$ and $\Delta E_{\nu}$ shows consistency with the
breakdown mechanism based on this picture too. In addition, our
experimental finding that there is a non-linear evolution of
dissipation boundary can possibly be attributed to Hall field
induced broadening of the extended state band.

\begin{figure}
\includegraphics[width=65mm]{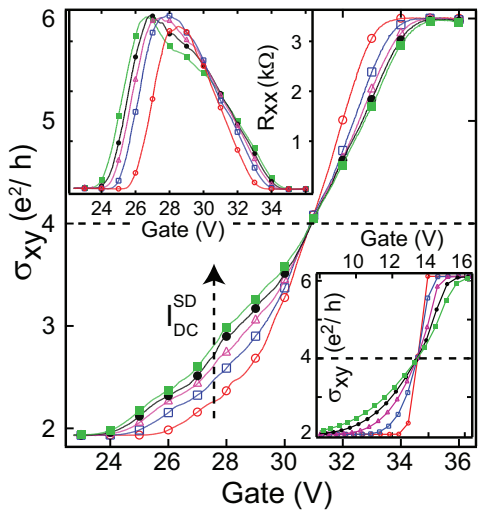}
\caption{\label{fig:figure3} (color online) Plot of $\sigma_{xy}$ as
a function of $V_g$ for $\nu=2$ to $\nu=6$ plateau transition at $T
$= 300 mK and $B$ = 10 T for different values of currents starting
from 0.75 $\mu A$ with an increment of 1.5 $\mu A$. The invariant
point at $\nu=4$ is clearly seen. The top-left inset shows the plot
of $R_{xx}$ as a function of $V_g$ for the same transition at the
same values of current as indicated in the main plot. The
bottom-right inset shows the calculated values of $\sigma_{xy}$ as a
function of the $V_g$. The invariant point at $\nu=4$ is also
clearly seen. }
\end{figure}

To further explore the effect of transverse electric field, due to
the high current density, we look at the plateau to plateau
transition in transverse conductance ($\sigma_{xy}$).
Fig.\ref{fig:figure3} shows the $\nu=2$ to $\nu=6$ plateau
transition at $T$ = 300 mK and $B$ = 10 T for different values of
current. As we increase the injected current, the transition width
starts to increase as well. Interestingly, in the transition region,
all the curves intersect at the filling factor 4 and $R_{xx}$ shows
a small suppression in peak resistance around the same gate voltage
($R_{xx}$ is shown in top-left inset). Such an invariant point
indicates that as we increase the current, the center of the
electric field induced broadened extended state band does not move
with the current. For $\nu=2$ to $\nu=6$ plateau transition, the
Fermi level crosses the four fold degenerate $n=1$ LL. It has been
shown that at very high magnetic field, spin degeneracy can be
lifted \cite{LiftingLL} giving rise to an additional plateau at
$\nu=4$. We speculate the current invariant point at $\nu=4$ and
suppression of $R_{xx}$ at the same time as a precursor of Zeeman
splitting. To understand our data quantitatively, we carried out
numerical calculations based on the injection model of QHE in
graphene\cite{KramerGraphene}. This model gives transverse
conductance from the calculation of local density of states. The
bottom-right inset in Fig.\ref{fig:figure3} shows the conductance
curves calculated for different values of the injected current. This
model accurately describes the position of the current invariant
point but fails to explain the width of the transition region. One
possible reason for the failure of this model could be the
assumption that all states are extended. Further detailed analysis
is needed to take into account the effect of disorder.

 In summary, we have studied the non-equilibrium breakdown
of the quantum Hall state in graphene. We find that the
dissipationless QH state can be suppressed due to a high current
density, and the corresponding critical current decreases with $|\nu
|$. The correlation between $V_{Hall}$ and $\Delta E_{\nu}$ is
consistent with the disorder-induced broadening of LLs and
inhomogeneous charge distribution. The value of $V_{Hall}$ at
breakdown gives an idea about the activation energy. Scanned probe
based measurements on cleaner samples are likely to observe the
electric field induced ``collapse" ($\beta\geq1$) of LLs. We also
see a current invariant point in the plateau to plateau transition
and suppression in longitudinal resistance at higher current, which
can possibly be a sign of spin-degeneracy breakdown.

\begin{acknowledgments}
The authors would like to thank Hari Solanki, Sajal Dhara, Shamashis
Sengupta, Prita Pant and Arnab Bhattacharya for their help.
% during the experiments and ******
%for helping in the preparation of this manuscript.
This work was
supported by the Government of India.
\end{acknowledgments}


\begin{thebibliography}{99}
\bibitem{noteaboutnu}

Nominally, the filling factor is defined as the ratio of number of
charge carriers to the number of magnetic flux quanta penetrating
the sample and therefore is positive. However, in the context of QHE
in graphene a negative filling factor is used to distinguish the
filling factor of holes from that of electrons
(cf.\cite{firstgraphene,grapheneRMP}).
\bibitem{absenceofnu10}
The $\nu=10$ is not studied as the gate voltage required to study it
can lead to a dielectric breakdown of the gate-dielectric SiO$_2$.
\end{thebibliography}
\end{document}